\documentclass[journal]{IEEEtran}
\usepackage{graphicx}
\usepackage{subfigure}
\ifCLASSINFOpdf

\else

\fi

\begin{document}
\title{Identifying Design Requirements for\\ Wireless Routing Link Metrics}

\author{\IEEEauthorblockN{Nadeem Javaid$^{\dag,\sharp}$, Muti Ullah$^{\ddag}$, Karim
    Djouani$^{\dag,\S}$\\\vspace{0.4cm}}
    \IEEEauthorblockA{
    $^{\dag}$LISSI, Universit\'e Paris-Est Cr\'eteil (UPEC), France. \{nadeem.javaid,djouani@univ-paris12.fr\}}\\
    $^{\sharp}$Dept. of Electrical Engg., COMSATS, Islamabad, Pakistan. \{nadeemjavaid@comsats.edu.pk\}\\
    $^{\ddag}$ICIT, Gomal University, D.I.Khan, Pakistan. \{matiullahns@gmail.com\}\\
    $^{\S}$F'SATI, Pretoria, South Africa. \{djouanik@tut.ac.za\}
     }

\vspace{100pt}

%\author{\IEEEauthorblockN{Nadeem Javaid$^{\dag}$, Khaled Dridi$^{\dag}$,MutiUllah Qureshi$^{\ddag}$, Karim
%    Djouani$^{\dag,\S}$\\\vspace{0.4cm}}
%    \IEEEauthorblockA{ $^{\dag}$LISSI, Universit\'e Paris-Est Cr\'eteil (UPEC), France. \{nadeem.javaid,djouani@univ-paris12.fr\}}\\
%    $^{\ddag}$ICIT, Gomal University, D.I.Khan, Pakistan. mutiullah@ieee.org\\
%    $^{\S}$F'SATI, Pretoria, South Africa. djouanik@tut.ac.za
%     }
%
%% make the title area

\maketitle

\begin{abstract}
%\boldmath
In this paper, we identify and analyze the requirements to design a new routing link metric for wireless multi-hop networks. Considering these requirements, when a link metric is proposed, then both the design and implementation of the link metric with a routing protocol become easy. Secondly, the underlying network issues can easily be tackled. Thirdly, an appreciable performance of the network is guaranteed. Along with the existing implementation of three link metrics Expected Transmission Count (ETX), Minimum Delay (MD), and Minimum Loss (ML), we implement inverse ETX; \textit{\textbf{invETX}} with Optimized Link State Routing (OLSR) using NS-2.34. The simulation results show that how the computational burden of a metric degrades the performance of the respective protocol and how a metric has to trade-off between different performance parameters.
\end{abstract}


\IEEEpeerreviewmaketitle

\section{Introduction}

\IEEEPARstart{P}{ERFORMANCE} of Wireless Multi-hop Networks (WMhNs) depends upon the efficiency of the routing protocol operating it and the most important component of a routing protocol is 'routing link metric'. Because, a link metric first considers the quality routes then decides the best end-to-end path. The link metric plays a key role to achieve the desired performance of the underlying network by making the routing protocol: fast enough to adopt topological changes, light-weight to minimally use the resources of nodes, intelligent to select the fastest path from source to destination among the available paths and capable to enable the nodes to have a comprehensive idea about the topology.

	Considering the demands of a wireless multi-hop network from its operating protocol and the factors influencing its performance, a metric is supposed to fulfill certain requirements. An efficiently designed routing metric can better help a routing protocol to achieve appreciable performance from the underlying network by dealing with these issues. In this work, we, therefore, identify the characteristics that must be taken into account while designing a routing link metric. It is worth stating that it is impossible to implement all mentioned characteristics in a single metric. Rather they provide guidelines that might be helpful to design a link metric. For instance, we have proposed and validated a new routing metric, Interference and Bandwidth Adjusted ETX (IBETX) in Wireless Multi-hop Networks \cite{0}, that considers link asymmetry, bandwidth and interference of the wireless links in the same contention domain. By simulation results we have demonstrated that the computational overhead produced by a routing metric may degrade the performance of the protocol. The issues that influence a wireless network, if efficiently tackled, they become the characteristics of the newly developed protocol.

\section{Related Work and Motivation}
After analyzing reactive and proactive protocols, Yang \textit{et al.} \cite{1} proposed that the proactive protocols that implement the hop-by-hop routing technique, as Destination-Sequenced Distance Vector (DSDV) \cite{2} and Optimized Link State Routing (OLSR) \cite{3} protocols are the best choice for mesh networks. They have also inspected the design requirements for routing link metrics for the mesh networks and related them to the routing techniques and routing protocols. In the chapter, four design requirements for link metrics; stability, minimum hop count, polynomial complexity of routing algorithm and loop-freeness have been suggested. However, the focus has only been on the mesh networks. Secondly, all the work is merely restricted to these four requirements. There are several other requirements that may help to achieve global optimization. For example, 'computational overhead' that might be outcome of the mathematical complexity introduced in the link metric or an attempt to design a multi-dimensional metric to tackle multiple issues simultaneously.

Das \textit{et al.} in \cite{4}, have discussed the dynamics of the well known metrics: Expected Transmission Count (ETX) \cite{5}, Expected Transmission Time (ETT) \cite{6} and Link Bandwidth \cite{7}, in real test beds. Across various hardware platforms and changing network environments, they tested two requirements: stability and sensitivity for some existing routing link metrics. Authors have also discussed the dynamics of the above mentioned metrics and tested their performance on the test beds for the above stated requirements. Anyhow, both the design issues of the link metrics and their design requirements are yet to be analysed.
	
In \cite{8}, Yaling \textit{et al.} systematically analyzed the impact of working of wireless routing link metrics on the performance of routing protocols. They related the characteristics of routing metrics to reactive and proactive protocols. They have presented the ways by which the mathematical properties of the weights given to the paths affect the performance of routing protocols. They proposed and discussed three operational requirements: optimality, consistency and loop-freeness. However,
these properties do not cover all design requirements; for example, computational overhead, a metric can produce and the performance trade-offs a metric has to make among different network performance factors. For example, a routing protocol achieves higher throughput values at the cost of end-to-end delay or routing overhead. So, instead of generalizing the design requirements, we have pointed-out and analyzed almost all possible design requirements.

\section{Factors Influencing WMhNs}
The factors affecting the wireless networks help to have an idea about the problems they have to face. Along with other protocols that operate a
network, routing protocols play a significant role in the performance of wireless multi-hop networks. So, in this section, we state and discuss some general issues regarding wireless networks that will provide a ground to discuss the requirements for designing a routing metric.

\textit{(A)} In wireless networks, generally the link quality considerably varies in different periods of time. The reasons may be: some mobile nodes are moving randomly, some go-out of range, some intentionally cut-off the ongoing communication, some die-out due to battery and so on. The respective routing protocols must be able to dynamically cop with the situation.

\textit{(B)} Usually, the behavior of channels varies in links and then in complete paths from source to destination. In the case of Quality of Service (QoS) routing, the the link creating bottle neck for performance must be given attention. Similarly, change in the quality of one link affects the others, as in the case of intra-flow and inter-flow interferences but not in the case of (minimum) hop count.

\textit{(C)} Upper layer protocols are affected by the choice of a particular link metric at the lower layers \cite{4}.

\textit{(D)} The selection for a particular flow on a particular channel is not random in the case of multiple flows on multiple channels.

\textit{(E)} The wireless multi-hop networks in which each node is equipped with a single radio interface and all radio interfaces operate on the same frequency channel, often suffer low channel utilization and poor system throughput.

After discussing the behavior of wireless networks, it would be appreciable to discuss and analyze the design requirements for routing link metrics.

\section{Design Requirements for Routing Metrics}
	Heretofore, several routing protocols either have been designed from scratch or optimized to improve the performance of a particular wireless network. A routing protocol is responsible to choose the best paths from source to destination. This decision is based upon the information provided by link metric. Therefore, primary emphasis has been given to propose new link metrics of different varieties; a single metric, a single mixed metric, a single compound metric, multiple metrics and a composite metric are few examples that have been designed and implemented with the existing protocols \cite{9}. Thus, while designing a link metric for a routing protocol, following design requirements must be taken into account.

%\begin{table}[h]
%\caption {LOAD BALANCING PROTOCOLS WITH RESPECTIVE TECHNIQUES}
%\vspace{-0.3cm}
%\begin {center}
%\begin{tabular}{|c|c|}
%\hline
%\textbf{Routing protocol} &\textbf{Load balancing Techniques }\\
%\hline
%
%Load Balanced Ad-hoc Routing & Measuring activity of \\
%
% (LBAR) \cite{12}&  a node participating in\\
%
%&the communication\\
%\hline
%
%&Counting the total number of \\
%Load Sensitive Routing(LSR) \cite{13} &packets both at the queue of mobile \\
% &node and neighboring node \\
%\hline
%
%Dynamic Load Aware Routing  &Measuring the routing over head at  \\
% (DLAR) \cite{14}&the intermediate node \\
%\hline
%
%Simple Load-balancing Ad-hoc  &Measuring the forwarding load of \\
% Routing (SLAR) \cite{15}&the mobile nodes\\
%\hline
%
%Ad-hoc On-demand Distance  &\\
%Vector Routing with Load   &Analyzing the balance index\\
% Balancing (LB-AODV) \cite{16}&\\
%\hline
%
% &- Measuring hop counts and traffic \\
%Load Aware Routing in Ad-hoc&loads   for TCP source\\
%networks (LARA) \cite{17}&- Measuring level of contention for \\
% &non TCP sources\\
%\hline
%
%Simple Load-balancing Approach &Measuring the traffic load at the \\
% (SLA) \cite{18}&mobile node\\
%\hline
%
%Delay-based Load-Aware &Measuring the hop count \\
% On-demand Routing (D-LAOR)&and end-to-end delay\\
%  protocol \cite{19}&\\
%
%  \hline
%
%\end{tabular}
%\end{center}
%\end{table}

\vspace{-0.3cm}
\subsection{Minimizing hop-count or path length}
This is first of the several canonical design requirements, a link metric is supposed to fulfill that has a goal to route packets through minimum weight paths. Often a longer path increases the end-to-end delay and reduces the throughput of a path. So, the respective metric must prefer a path with minimum length over it. This design requirement is implicitly or explicitly attempted by almost all of the existing link metrics. For instance, \textit{ETX} achieves maximum throughput by minimizing the number of transmissions and thus raises a network throughput.
Minimum Loss \textit{(ML)} \cite{10} selects the paths with minimum loss rates or higher probabilities of successful transmissions. Now, if all links in some end-to-end paths have the same probabilities of success, then qualities of the paths becomes dependent on the number of hops. \textit{ML} has been implemented with OLSR that prefers minimum hop path in this case. Hop count is the most widely used metric in MANET routing protocols \cite{11}, as all of the RFC's prefer to use hop count as a routing metric for the sake of simplicity and least computational overhead.

\vspace{-0.3cm}
\subsection{Balancing traffic load}
To achieve appreciable throughput, the respective metric can be designed to ensure that no node or link is disproportionately used by minimizing the difference between the maximum and minimum traffic load over the nodes or links.

When a link becomes over-utilized and causes congestion, the link metric can choose to divert the traffic from the congested path or overloaded nodes to the underloaded or idle ones to ease the burden. 

\vspace{-0.3cm}
\subsection{Minimizing delay}
A network path is preferred over the others because of its minimum delay. It is worth noting that if intra-flow and inter-flow interferences, queuing delays, and link capacity are not taken into consideration, then delay minimization often ends up being equivalent to path length or hop-count minimization.

\vspace{-0.3cm}
\subsection{Maximizing data delivery/aggregating bandwidth }
Maximizing the probability of data delivery, minimizing the probability of data loss, minimizing the packet loss ratio, maximizing the packet delivery fraction, maximizing the individual path throughput, increasing the network capacity, are the same and utmost important features, a wireless routing protocol is expected to implement. So, in wireless networks, the attempt has always been to choose an end-to-end high capacity path. A protocol can achieve maximum throughput:

\textit{(a)} directly by maximizing the data flows,

\textit{(b)} indirectly by minimizing interference or retransmissions,

\textit{(c)} allowing the multiple rates to coexist in a network, where a higher channel rate is used over each link. It is possible if more packets can be delivered in the same period with the consideration of packet loss rates \cite{20}, data can be splitted to the same destination into multiple streams, each routed through a different path.  End-to-end delay may also be reduced as a direct result of larger bandwidth.

\vspace{-0.3cm}
\subsection{Minimizing energy consumption}
Energy consumption is a major issue in all types of wireless networks where the battery lifetime constrains the autonomy of network nodes. A protocol, if chooses path with an unreliable link, it would probably produce longer delay due to higher retransmission rates, that ultimately results in raise in energy consumption (along with computational processing overhead of aggressive control packets). For energy saving, most of the work focuses on the communication protocol design. For example, the routing protocol ZigBee \cite{21} uses a modified AODV to be used by low-power devices. By adapting transmission power to the workload, Real-time Power-Aware Routing (RPAR) protocol \cite{22} reduces communications delays.

\vspace{-0.3cm}
\subsection{Minimizing channel/interface switching}
Both in single-hop and multi-hop wireless networks, for the maximum utilization of available bandwidth, one way is to use as many channels as possible depending upon the sophistication of the technology. In this case, the different data flows are to be switched on different channels, resulting in some delay. So, the phenomenon may be given attention by the respective metric.

	When using multiple channels, two adjacent nodes can communicate with each other only if they have at least one interface on a common channel. So, it may be necessary to periodically switch interfaces from one channel to another with the production of a delay. In \cite{23}, Vaidya \textit{et al.} used an interface assignment strategy that keeps one interface fixed on a specific channel, while other interfaces can be switched among the remaining channels, when necessary.

\vspace{-0.3cm}
\subsection{Minimizing the Computational overhead}
While designing a routing metric, necessary computations should be considered that must not consume memory, processing capability and the most important; battery power. For example, we discuss the case of three widely used routing link metrics for wireless routing protocols: \textit{ETX}, its  inverse, say, \textit{invETX} and \textit{ML}.

For an end-to-end path, $P_{e2e}$, these metrics are  expressed by the following equations:

\small
\begin{eqnarray}
 ETX_{P_{e2e}}=\sum_{l \in P_{e2e}}^{}\frac{1}{(d_f^{(l)}\times d_r^{(l)})}
\end{eqnarray}
\normalsize

\small
\begin{eqnarray}
 invETX_{P_{e2e}}=\sum_{l \in P_{e2e}}^{}{(d_f^{(l)}\times d_r^{(l)})}
\end{eqnarray}
\normalsize

\small
\begin{eqnarray}
  ML_{P_{e2e}}=\prod_{l \in P_{e2e}}^{}(d_f^{(l)}\times d_r^{(l)})
\end{eqnarray}
\normalsize

\vspace{-0.3cm}
\subsection{Minimizing interference}
Bandwidth of a wireless link is shared among neighboring nodes, so, the contending nodes have to suffer from the inter-flow interference. The channels on the same link are always being disturbed from the intra-flow interference. Both intra-flow and inter-flow interferences may result in bandwidth starvation for some nodes as they may always find the available channels busy. Hence, both of the diversity of channel assignments and the link capacity possibly need to be captured by the link metric, as Yang \textit{et al.} have presented in their work.

Where $(d_f^{(l)}\times d_r^{(l)})$ is the probability of success for delivery of probe packets (134 $bytes$ each) on the link $l$ on $P_{e2e}$ from source to destination (forward direction) and from destination to source (reverse direction).

Regarding the computational complexity, all of the three metrics have to calculate the equal number of products $(d_f^{(l)}\times d_r^{(l)})$ for the same number of links. But $ETX$ has to suffer from more computational overhead (inverse and sum of $n$ products) than $ML$ (multiplication of $n$ products only). Similarly, $ML$ generates more computational overhead than $invETX$. As a result, $invETX$ achieves higher throughputs than $ML$ and $ETX$. Similarly, $ML$ performs better than $ETX$. The computational overheads generated by the three metrics have been shown in Fig. 1.a. Along with other implementation parameters, the amount of computational load generated by each metric influences its performance accordingly. This fact can be seen in Fig. 1.b, 1.c and 1.d. This overhead is directly proportional to the number of nodes/links.

\vspace{-0.3cm}
\subsection{Maximizing route stability}
Unlike wired networks, frequent topological changes in the wireless links may not only huge generate routing load but may also slow down the convergence of the respective routing protocol operating the network. The stability of the paths is found by the path characteristics that are captured by the routing metric that can be either load sensitive or topology-dependent \cite{1}. Former type of metrics assign a weight to a route according to the traffic load on the route. This weight may change frequently as the link break and establish. On the other hand, topology dependent metrics assign a weight to a path based on the topological properties of the path, such as the hop-count and link capacity of the path. Therefore, topology-dependent metrics are generally more stable, especially for static networks where the topology does not change frequently. Load-sensitive and topology-dependent metrics are best used with different types of routing protocols, since routing protocols have different levels of tolerance of path weight instability \cite{24}.

\vspace{-0.3cm}
\subsection{Maximizing fault tolerance/minimizing route sensitivity}
In the case of multi-path routing, the link metric can provide fault tolerance by having redundant information of the alternative paths. This reduces the probability that communication is disrupted in the case of link failure. To reduce the network load due to the redundancy, source coding can be employed with the aid of some sophisticated algorithms with compromising on the issue of reliability. Such type of raise in route resilience usually dependents upon the diversity, or disjointness like metrics for the available paths \cite{24}.

\vspace{-0.3cm}
\subsection{Avoiding short and long lived loops}
A metric can better help a routing algorithm to avoid forwarding loop (both short lived and long lived) to minimize the packet loss. Because selecting redundant links degrades the performance of the network due to more path lengths and consequently increased end-to-end delay. For example, Faheem \textit{et al.} \cite{25} have addressed the problem of transient mini-loop problem that takes place because of fisheye scoping in Fish Eye OLSR (OFLSR) protocol. They have provided a potential solution that enables the routers to calculate "safe" scope for a particular topology for all updates. The minimum TTL value that eliminates mini-loops, is calculated in distributed fashion by all mesh routers in advance at the "scope" boundary. Independent of the scale of network, keeping efficiency of the algorithm as before, the authors improved the safety of OFLSR.

\vspace{-0.3cm}
\subsection{Considering performance trade-offs }
Generally, a protocol achieves higher throughput values at the cost of increased end-to-end delay in the case of static networks. Whereas, in mobile networks, the frequent link beaks cause more routing overhead to obtain better throughput from the network. To discuss such type of trade-offs, we have set-up a simulation scenario that is discussed in the following section.

\begin{figure}[h]
  \centering
 \subfigure[Computational Overhead of ETX, ML, InvEX]{\includegraphics[height=3 cm,width=4.3 cm]{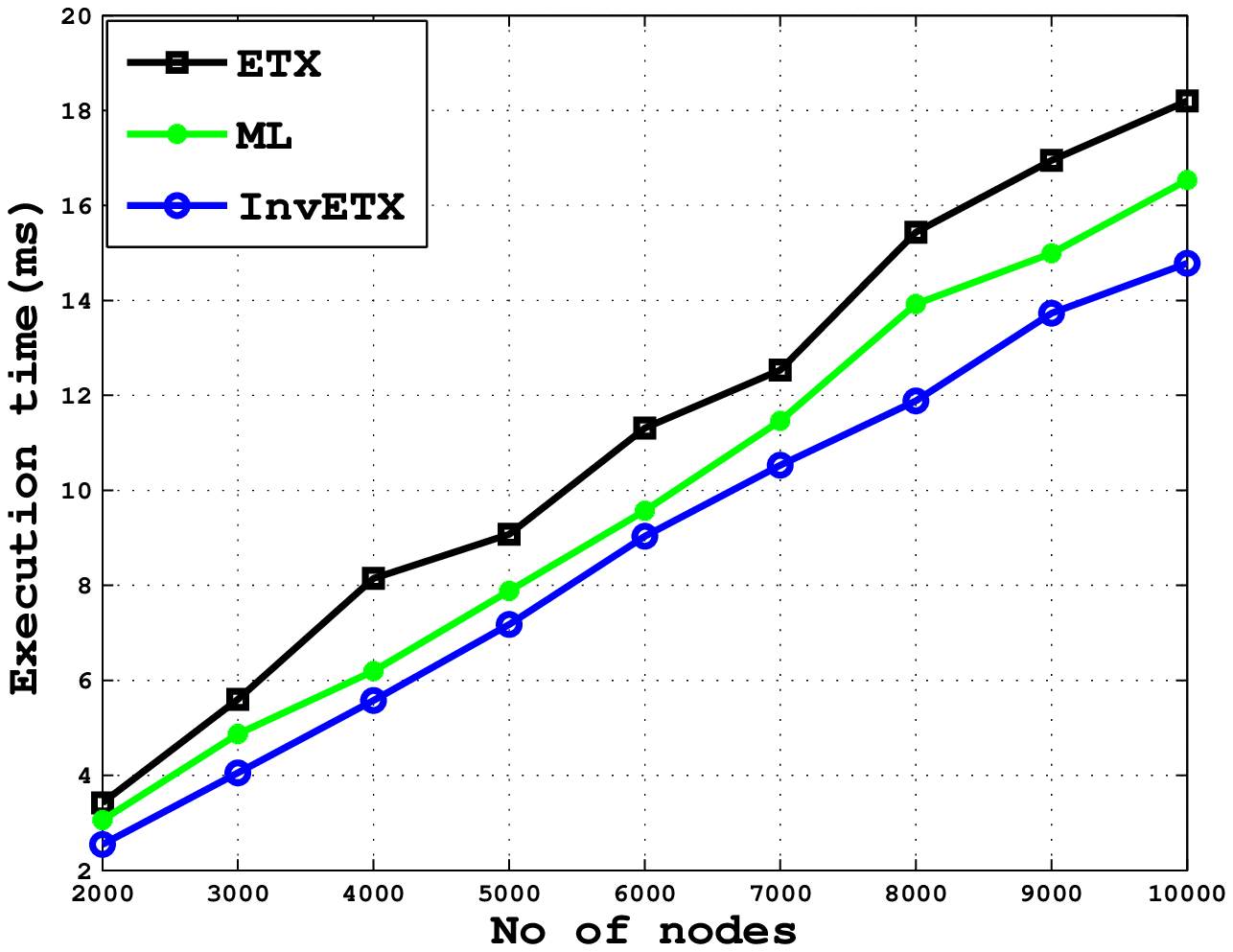}}
 \subfigure[Throughput of OLSR with 4 metrics]{\includegraphics[height=3  cm,width=4.3 cm]{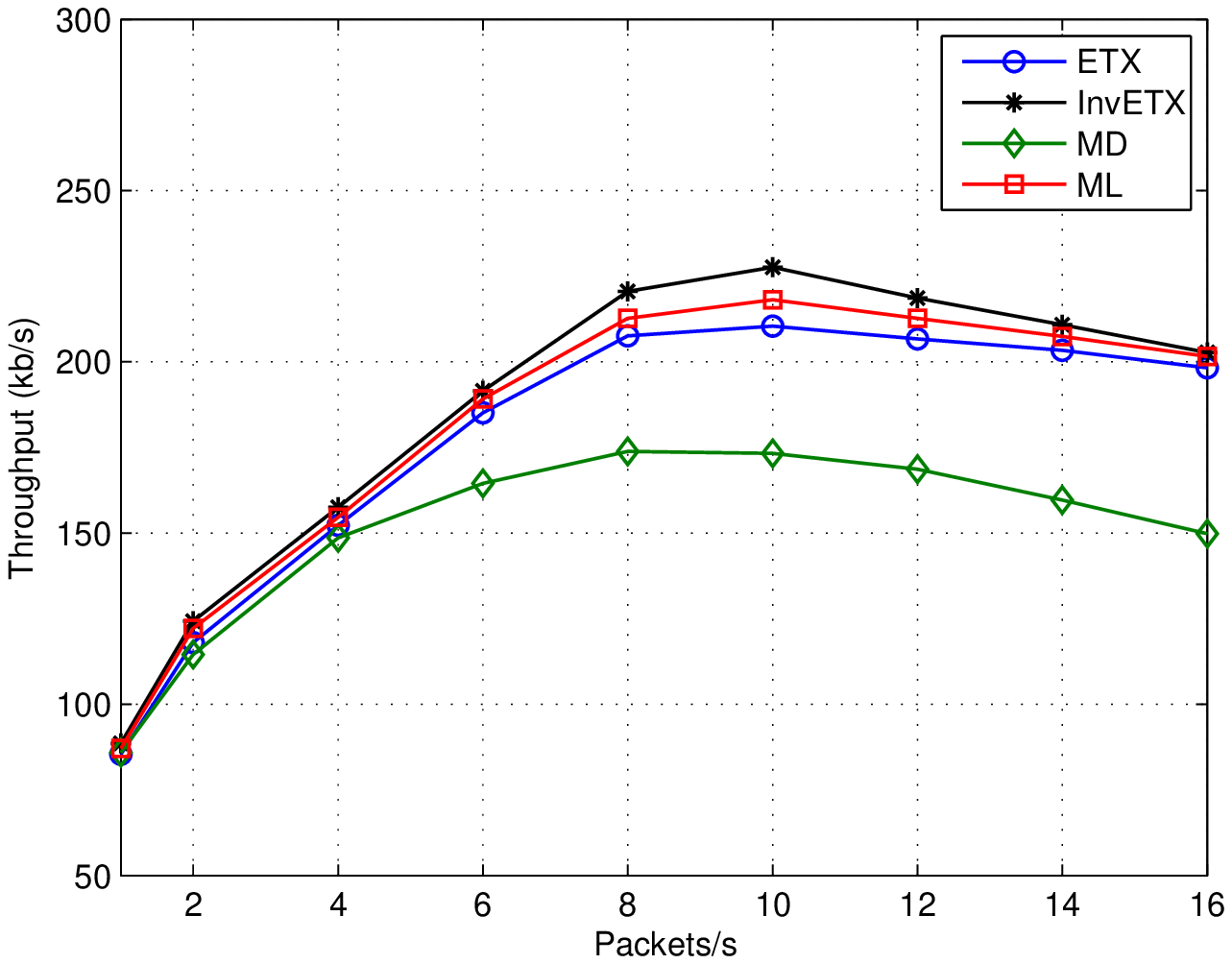}}
  \subfigure[E2ED of OLSR with 4 metrics]{\includegraphics[height=3 cm,width=4.3 cm]{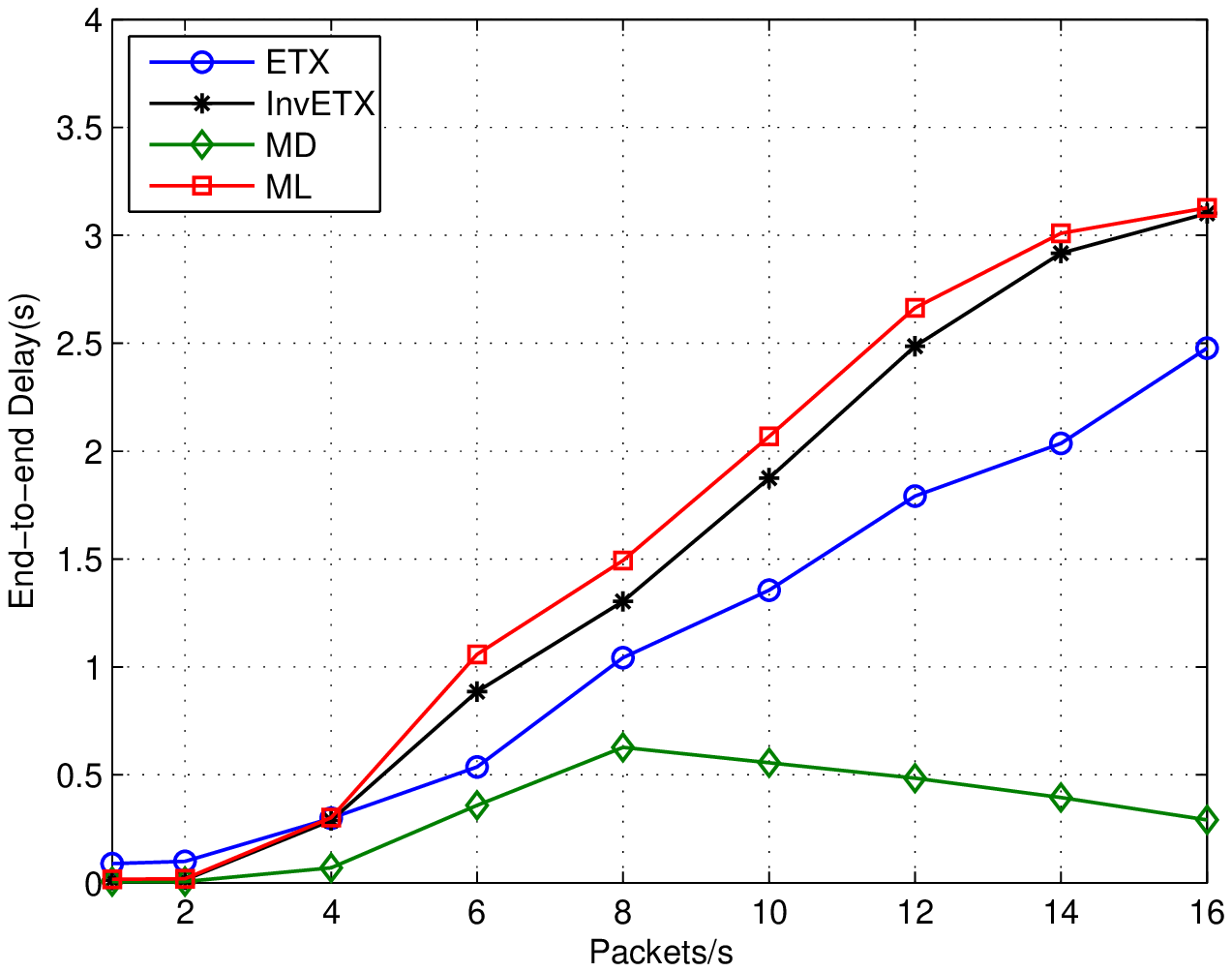}}
 \subfigure[NRL of OLSR with 4 metrics]{\includegraphics[height=3  cm,width=4.3 cm]{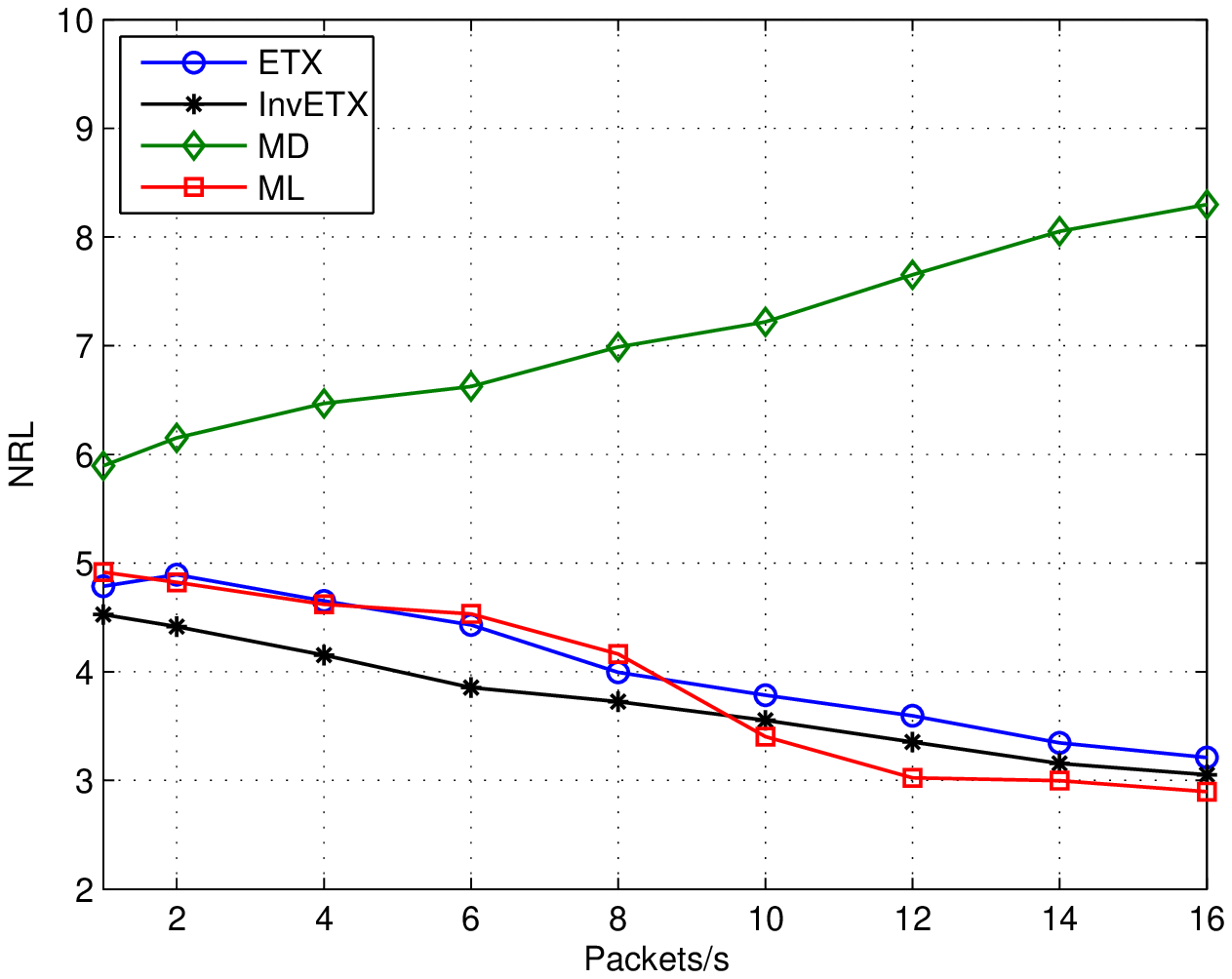}}
  %\caption{FSR performance analysis with varying packet rates and scalabilities}
\end{figure}
\vspace{-0.3cm}
\hspace{-0.3cm}\small Figure.1. Computational Overhead, Throughput, Delay, Routing load by Metrics

\section{Simulations}

We use the implementation of \textit{ETX}, Minimum Delay \textit{(MD)} \cite{26}, and \textit{ML} \cite{10} with OLSR \cite{9} in NS2-2.34. Then we implement the fourth metric, \textit{invETX}, as expressed by eq. (2). In the area of 1000$m$ x 1000$m$, 50 nodes are placed randomly to form a static network. Constant Bit Rate (CBR) traffic is randomly generated by 20 source-destination pairs with packet size of 64$bytes$. Each simulation is performed for five different topologies for 900$s$ each. Then the average of five different values of each performance parameter is used to plot the graphs. To observe the performance of OLSR with four metrics, we randomly generated the data traffic with number of packets from 1 to 16 per second.

To better understand the performance trade-offs, we take an example of the static wireless multi-hop networks that have two major issues; bandwidth and end-to-end delay. In this type of networks, the proactive protocols are preferred due to stability, like, OLSR, instead of the reactive ones that are suitable for the environments where topology changes frequently due to mobility. Moreover hop-by-hop routing technique helps OLSR to handle aggressive overhead as compared to source routing. Using the Multi-point Relays (MPRs) selection along with proactive nature, OLSR achieves minimum delay. In the following subsections, we discuss the performance parameters; throughput, End-to-End Delay \textit{(E2ED)}, and Normalized Routing Load \textit{(NRL)}.\newline
\textbf{Throughput}
In static networks, with varying data traffic rates, \textit{OLSR-MD} produces lowest throughput as compared to \textit{OLSR-ETX}/\textit{OLSR-invETX} and \textit{OLSR-ML}. Moreover, in medium and high network loads, there are more drop rates as compared to small load in the case of \textit{MD} metric.

This is due to the one-way delays that are used to compute the \textit{MD} routing metric with small probe packets before setting up the routing topology and not considering the traffic characteristics. It may thus happen that, if no other traffic is present in the network, the probes sent on a link experience very small delays, but larger data packets may experience the higher delay or retransmission due to congestion. Thus, \textit{OLSR-MD} is not suitable for the static networks with high traffic load, as, it degrades the network performance by achieving less throughput values. The \textit{OLSR-ML} in medium and high network loads produces higher throughput values because \textit{ML} attains the less drop ratios as compared to \textit{ETX}. Moreover, in \textit{ML} the paths with minimum loss rates or higher probabilities of successful (re)transmissions lead to high data delivery rates, with an additional advantage of more stable end-to-end paths and less drop rates.

\textit{OLSR-MD} uses the Ad-hoc packet technique to measure the one-way delay. Then proactive delay assurance approach is used to measure \textit{MD} metric. The minimum delay metric performs best in terms of average packet loss probability. In Fig. 1.c, \textit{OLSR-MD}'s delay is showing  the lowest values among other metrics. This is due to the route selection decision based on delay of ad-hoc probes. While \textit{OLSR-ETX} and \textit{OLSR-ML} produce increasing value of delay, when traffic increases. The very first reason is that both metrics have no mechanism to calculate the round trip, unlike \textit{MD} metric. Meanwhile, in \textit{ML}, selection of longer routes with high probability of successful transmission augments the delay as compared to \textit{ETX}.\newline
\textbf{NRL}
\textit{OLSR-MD} suffered from the highest routing loads. As, ad-hoc probes are used to measure the metric values and are sent periodically along with TC and HELLO messages. On the other hand, \textit{OLSR-ETX} and \textit{OLSR-ML} calculate the probabilities for the metric from the values obtained from the enhanced HELLO messages.

OLSR uses HELLO and TC messages to calculate the routing table and these messages are sent periodically. The delivery ratios are measured using modified OLSR HELLO packets that are sent every $t$ seconds ($t=2$, by default).

Each node calculates the number of HELLO messages received in a $w$ second period ($w=20$, by default) and divides it by the number of HELLO messages that should have been received in the same period (10s, by default). Each modified HELLO packet notifies the number of HELLO messages received by the neighbor during the last $w$ seconds, in order to allow each neighbor to calculate the reverse delivery ratio. The worse the link quality, the higher the \textit{ETX} link value. A link is perfect if the \textit{ETX} value is 1 and its packet delivery fraction is also $1$, i.e., no packet loss. On the other hand, if in $w seconds$ period a node has not received any HELLO message then \textit{ETX} is set to 0 and the link is not considered for routing due to 100\% loss ratio. Thus, due to no extra overhead to measure the metric \textit{OLSR-ETX}/\textit{OLSR-invETX} and \textit{OLSR-ML} have to suffer from low routing load as compared to \textit{OLSR-MD}.

The ad-hoc probe packets are sent by \textit{MD} to accurately measure the one-way delay. Thus, low latency is achieved by selecting the path with less Round Trip Time (RTT). On the other hand, these ad-hoc probes cause routing overhead in a network and decrease the throughput when data load is high in a static network.

In static networks, to measure an accurate link with less routing load is a necessary condition. The delay cost due to increase in the number of intermediate hops is paid to achieve throughput by \textit{OLSR-ML}. As \textit{ML} selects those paths which possess less loss rates, therefore, a longer path with high successful delivery is preferred. Thus the product of the link probabilities selection decreases the drop rates and increase the RTT.

\textit{OLSR-ETX} uses the same mechanism to measure the link quality as that of \textit{OLSR-ML}, i.e., modified HELLO messages. But summing up the individual probabilities and preference of the shortest path reduces the delay of \textit{ETX} as compared to \textit{ML}. Thus, a slow link preference results more drop rates of \textit{OLSR-ETX} as compared to \textit{OLSR-ML}.

This sort of trade-off is common in routing protocols. While designing a link metric, if demands of the underlying network are taken into consideration then it becomes easy to decide that among which performance parameters, trade-off(s) should be made. For example, \textit{ML} and \textit{ETX} achieve higher throughput values than \textit{MD}, as shown in Fig. 1.b, whereas \textit{MD} remarkably achieves less end-to-end delay than \textit{ML} and \textit{ETX} that is depicted in Fig. 1.c.

In Table 1, we provide a list of routing link metrics and routing algorithms that have taken into account some of the design requirements suggested in this chapter.\newline\newline
\small Table.1. Metrics implementing different Design Requirements\normalsize
\vspace{-0.5cm}
\begin{table}[!h]
%\caption {Metrics and Algorithms implementing different Design Requirements}
\begin {center}
\begin{tabular}{|c|c|}
\hline
\textbf{Design Requirement} & \textbf{Metric/Algorithm}\\
\hline

Minimizing hop count  &Hop count \cite{11}\\
or path length&\\
\hline

Minimizing delay&Per hop RTT \cite{6} \\
\hline

Minimizing packet loss ratio  &Interference clique   \\
 &transmissions \cite{27} \\
\hline

Balancing traffic load  &MIC \cite{11} \\
 \hline

Maximizing the probability  &Per hop PktPair \cite{6}, ML \cite{10}\\
of data delivery   &\\
\hline

 Maximizing path capacity&Network characterization  \\
&with MCMR \cite{28}\\
\hline

Aggregating bandwidth/ &Multipath routing scheme for  \\
 maximizing  fault tolerance&wireless ad-hoc networks\cite{10}\\
\hline

Maximizing individual &ETX \cite{5}, \cite{11}\\
path throughput&\\
\hline

Max. throughput of individual link &ETX \cite{5}, \cite{11}\\
\hline

Maximizing network throughput &ETX \cite{5}, \cite{11}\\
\hline

Max. throughput of individual link &ETX \cite{5}, \cite{11}, Per hop RTT \cite{11}\\
\hline

Minimizing interference &iAWARE \cite{11}\\
\hline

Minimizing channel switching&MIC \cite{11}, WCETT \cite{11}\\
\hline

Minimizing interface switching&MCR Protocol \cite{11}\\
\hline

Maximizing route stability&Link affinity metric\cite{29}\\
\hline

Minimizing energy consumption&MTPR \cite{30}, MBCR \cite{31}\\
\hline

Avoiding routing loops&Loop avoidance for Fish-Eye   \\
  &OLSR in Sparse WMN's \cite{25}\\
\hline

Minimizing computational overhead&ML \cite{10}\\
\hline

\end{tabular}
\end{center}
\end{table}

\vspace{-0.3cm}

\vspace{-0.3cm}
\section{Conclusion and Future Work}
In this work, we present a comprehensive study on the design requirements for routing link metrics. We select Expected Transmission Count (ETX), Minimum Delay (MD), Minimum Loss (ML) and our proposed meric; Inverse ETX (invETX) with OLSR. We discuss several possible issues regarding wireless networks that can better help in designing a link metric. The ambition of a high throughput network can only be achieved by targeting a concrete compatibility of the underlying wireless network, the routing protocol operating it, and routing metric; heart of a routing protocol. Depending upon the most demanding features of the networks, different routing protocols impose different costs of 'message overhead' and 'management complexity'. These costs help to understand that which type of routing protocol is well suitable for which kind of underlying wireless network and then which routing link metric is appropriate for which routing protocol. In future, we are interested in an analytical study of such kind of compatibility.

%\ifCLASSOPTIONcaptionsoff
%  \newpage
%\fi

\end{document}